# Interfacial Magnetism in Complex Oxide Heterostructures Probed by Neutrons and X-rays

Yaohua Liu[1] and Xianglin Ke[2]

[1.] Quantum Condensed Matter Division, Oak Ridge National Laboratory, Oak Ridge, TN, USA; email: liuyh@ornl.gov
[2.] Department of Physics and Astronomy, Michigan State University, East Lansing, MI, USA; email: ke@pa.msu.edu

Magnetic complex-oxide heterostructures are of keen interest because a wealth of phenomena at the interface of dissimilar materials can give rise to fundamentally new physics and potentially valuable functionalities. Altered magnetization, novel magnetic coupling and emergent interfacial magnetism at the epitaxial layered-oxide interfaces are under intensive investigation, which shapes our understanding on how to utilize those materials, particularly for spintronics. Neutron and x-ray based techniques have played a decisive role in characterizing interfacial magnetic structures and clarifying the underlying physics in this rapidly developing field. Here we review some recent experimental results, with an emphasis on those studied via polarized neutron reflectometery and polarized x-ray absorption spectroscopy. We conclude with some perspectives.

## 1. Introduction

Layered transition-metal oxides, in particular strongly correlated electron systems of perovskites, have attracted considerable attention in condensed matter and materials physics communities. For example, the family of colossal magnetoresistance manganites, based on $LaMnO_3$ and its derivatives, has served as a model system for studying the fundamental exchange interactions and the interplay between different degrees of freedom, including spin, charge, lattice and orbital [1-4]. A prominent feature of transition-metal complex oxides is the competing collective states with comparable energies, thus ground states with dramatically different physical properties can be reached via small perturbations, such as chemical doping, strain, and magnetic/electric fields. Enabled by advances in thin-film synthesis techniques, it is now feasible to create atomically sharp interfaces in epitaxial oxide heterostructures, opening new paths to engineer physical properties of complex oxides. This gives rise to new opportunities to realize systems in which one can introduce and explore compelling physical phenomena at the interface [5-8]. Several important effects can occur across oxide interfaces, including chemical reconstruction [9-11], charge transfer [12-14], structural coupling [15,16], and more interestingly, spin and orbital reconstructions [17-20].

Because of the strong coupling among different degrees of freedom, physical properties of transition metal oxides can be largely altered at the interface of interest, which can be crucial for functionalities but challenging to be predicted. One example is the suppressed magnetization at the interface when growing a high spin-polarization oxide on a nonmagnetic insulator [21-23], such as in magnetic tunnel junctions, which may produce inferior device performance [24,25]. There are lasting efforts to understand its origin and eliminate this effect [21,23,26]. Furthermore, novel magnetic coupling and emergent interfacial magnetization can have strongly affect functionalities, including exchange bias [18,19,27] and spin-dependent transport [20,28,29]. It is very important to use multiple complementary tools that are capable to characterize different aspects of oxide interfaces, in order to uncover the underlying physics of novel interfacial phenomena in strongly correlated systems. Profound understanding of the oxide interface physics will certainly contribute to utilize these intriguing properties for applications.

The conventional magnetometers, such as vibrating sample magnetometer (VSM) and superconducting quantum interference device (SQUID) magnetometer, are only sensitive to the total magnetization of samples. They lack spatial resolution and cannot discern between the contributions of different atoms in an alloy or multilayer, or between their orbital and spin moments, which often hinder the understanding of microscopic mechanisms of the novel magnetic phenomena. Moreover, the small sample quantity in many technologically relevant structures necessitates ultrasensitive probing techniques. For example, in order to measure weak magnetization signals from thin films grown on thick substrates, the contribution from the substrates is a serious concern. Therefore techniques with spatial resolution and/or element-specific capability are very advantageous. Neutron and x-ray based techniques, such as polarized neutron reflectometry (PNR) and polarized x-ray absorption spectroscopies (XAS), are traditionally among the most powerful techniques to study magnetic materials.

In this review, we focus on the neutron and x-ray work on heterostructures of perovskites and double perovskites, particularly those studied via PNR and polarized XAS. The first part is a brief introduction to these two techniques. We follow with several examples of interesting magnetic phenomena recently observed at complex oxide interfaces, which are closely related to device applications. Topics include suppressed interfacial magnetization of high spin polarization oxides [23,26], novel exchange coupling [27,30], emergent interfacial magnetization [18,19,31-33] and its effect on charge transport [20,28,31] and magnetization reversal [18,19]. This article is not intended to be a comprehensive review of complex oxide heterostructures, as emphasis is given to the studies performed using polarized neutrons and x-rays to probe the interfacial magnetism. Instead, we refer the readers to several recent literature for more information on the interfacial magnetic phenomena in oxide heterostructures [5-8,34,35]. We conclude with perspectives that are of high interest from the authors' viewpoints.

## 2. Polarized Neutron Reflectometry and X-Ray Absorption Spectroscopy

In this section, some basics on polarized neutron reflectometry and polarized x-ray absorption spectroscopy are given from an experimenters' view. With a brief introduction to the principle, we will discuss some practical aspects of the techniques, including information that can be readily obtained using these techniques, general sample requirements, and practices to minimize uncertainty and to improve the spatial and/or magnetization resolution. As magnetism probes, PNR is a vector magnetometer with a sub-nanometer spatial resolution, while polarized XAS is an element-specific magnetometer, which probes both ferro- and antiferromagnetism. As illustrated later in Sec. 3, such properties make these techniques sensitive to buried interfaces.

### 2.1 Polarized Neutron Reflectometry

Polarized neutron reflectometry is a nondestructive method to determine the chemical and magnetic structures in magnetic multilayers with sub-nanometer depth resolution. Information is obtained from the depth profiles of neutron scattering length density (SLD) via modeling. Typical experimental data are collected under the specular condition ($\theta_i = \theta_r$ with $\theta_i$ and $\theta_r$ being the angles of incident and reflected neutron beams relative to the film plane, respectively) in a magnetic field (See Fig. 1(a)). The intensity of the reflected beam is collected as a function of wavevector transfer, $Q_z = (4\pi \sin \theta)/\lambda$, where $\theta$ ($\theta = \theta_i = \theta_r$) is the incident angle and $\lambda$ is the neutron wavelength. $Q_z$ is changed via changing either the wavelength (~ 2-12 Å) or the incident angle (~ a few of degrees).

PNR is similar to x-ray reflectometry (XRR) that is widely used in thin film laboratories for thickness calibration. X-rays interact with the charge degree of freedom and the interactions are mostly from electrons. Thus specular XRR yields the depth profile of the electron density, which can be used to reconstruct the chemical structure. Neutrons interact with both nuclei and magnetization (because the neutron carries a spin 1/2), and the nuclear and magnetic neutron scattering lengths have comparable magnitudes. Polarized neutrons with spin parallel (spin-up, +) or antiparallel (spin-down, −) to the direction of the external magnetic field experience the same nuclear scattering potential, but opposite magnetic scattering potentials (Fig. 1(b)). From subsequent measurements with oppositely polarized neutron beams, these two contributions can be separated to obtain the depth profiles of both the chemical structure and the magnetization vector. The magnetic SLD is related to the magnetic moment density ($B$) by a constant c, where $c = 2.9109 \times 10^{-9}$ Å$^{-2}(\frac{kA}{m})^{-1}$. Polarization analysis of the specularly reflected beam provides information about the projection of the net magnetization vector onto the sample plane. One will typically collect four reflectivities, if the magnetization has a nonzero in-plane component that is perpendicular to the field, including two non-spin-flip (NSF) reflectivities, $R^{++}$ and $R^{--}$, and two spin-flip (SF) reflectivities, $R^{+-}$ and $R^{-+}$. For the NSF reflectivities, the neutron beam retains its original polarization after being scattered from the sample; while for the SF reflectivities, the reflected neutron beam flips its spin state. The NSF reflectivities provide information concerning the chemical composition and are sensitive to the component of the in-plane magnetization aligned along the field axis. The SF reflectivities are sensitive only to the component of the in-plane magnetization perpendicular to the field direction. Specular PNR is ideally suited to measure the nuclear and magnetization depth profiles across planar interfaces, because reflection occurs when scattering potential changes. Therefore, PNR is capable to probe a weak magnetization signal from thin films or interfaces with little influence from a substrate, which is a serious concern when a bulk magnetometery is used. In addition, off-specular scattering ($\theta_i \neq \theta_r$) originating from the in-plane correlations, contains in-plane neutron wavevector transfer thus probes the in-plane correlated roughness and magnetic stripe domains [27,36-38].

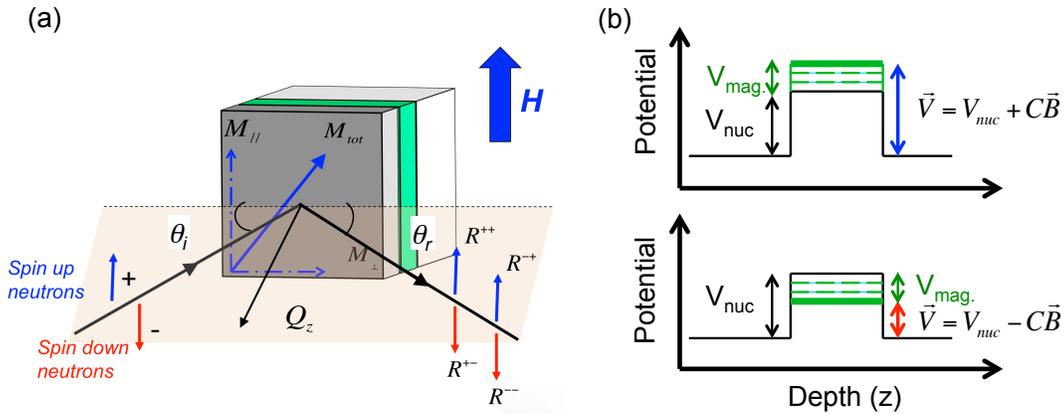

Figure 1. (a) Typical configuration of specular ($\theta_i = \theta_r$) PNR experiments, where neutron reflectivity is measured as a function of the wavevector transfer $Q_z$. The polarization direction of the neutron beam is either parallel or antiparallel to the applied field direction. Typically, four cross sections ($R^{++}$, $R^{+-}$, $R^{-+}$, $R^{--}$) will be measured to determine the depth profile of the averaged in-plane magnetization vector. The symbols of + and − label the spin polarization of the neutron beam being parallel and antiparallel to the laboratory field of reference $\boldsymbol{H}$, respectively. When off-specular ($\theta_i \neq \theta_r$) data are collected (not illustrated here), one is able to determine the in-plane correlated roughness and magnetic stripe domains. (b) Polarized neutrons with spin parallel (spin-up, +) or antiparallel (spin-down, −) to the direction of the external magnetic field experience the same nuclear scattering potential (scalar), but opposite magnetic scattering potentials (vector).

The spin polarization of a modern PNR reflectometer can be higher than 98%. However, polarization correction is still required when a high precision on the magnetization is needed, such as in cases of induced interfacial magnetization between two nonmagnetic oxides [39] and weak magnetization in multiferroic materials [40]. Reflectivities can be simulated based on the Parratt formalism [41]. A rough interface was modeled as a sequence of very thin slices, whose SLDs vary, followed by an error function so as to interpolate between adjacent layers. The effect of the instrumental resolution is typically handled by Gaussian convolution. If strong SF scattering occurs in a sufficiently high magnetic field (> 0.1 T), it is important to take into account the Zeeman effect when analyzing the data [42]. For more detailed description on the data reduction on the polarization correction and the Zeeman effect, see [42] and references therein.

There are some rules of thumb for thin-film sample requirement for PNR experiments, such as high in-plane homogeneity with a small surface roughness (root-mean-square < 1 nm) and a uniform thickness (~ 1 %) and a large area (~ 1 cm$^2$). Experiments on small samples are possible, but at the cost of the resolution and statistical quality of the data which scales as the area. Typical film thicknesses range from 10 nm to 100 nm. The practical magnetization and spatial sensitivity is of the order of 10 kA/m over a length scale of 0.5 nm. The feasibility to detect thinner magnetic layers and smaller magnetization variations depends on the maximum $Q$ value that has a good signal to noise ratio. The sample quality, the neutron flux, and the instrumental background can affect this. Note that reflectivity drops at least with a prefactor of $Q^{-4}$ at high $Q$, and the typical dynamic range of a modern PNR instrument is about 7 orders of amplitude. A common practice to enhance the interface sensitivity is to use superlattices [26,39]. Complementary approaches, such as x-ray reflectivity, are widely used to enhance the reliability of the model.

Because of the weak interaction between neutrons and most materials, one can implement sophisticated sample environments for PNR experiments, including high magnetic fields and low temperatures [39], mechanical gadgets to apply bending stress to substrates [43] and devices allowing light irradiation [44]. See Ref. [36,45] for more details on the PNR technique.

### 2.2 Polarized X-ray Absorption Spectroscopies

X-ray absorption spectroscopy provides information of the sample via probing the energy, angle, and/or polarization dependent absorption of x-rays. The absorption edges have characteristic energies for each element, which gives rise to the elemental sensitivity. Magnetic properties of transition metals are largely related to the $d$ orbitals, which are best probed by $L$-edge absorption. The $L$-edge XAS of the transition metal elements are dominated by two main peaks, separated by about tens of eV for 3$d$ ions, which are due to the excitations from the spin-orbit-split 2$p_{3/2}$ and 2$p_{1/2}$ core levels to empty 3$d$ valence states, respectively. The dipole selection rules determine which 2$p^5$3$d^{n+1}$ final state can be reached from a particular 2$p^6$3$d^n$ initial state and the transition probability. Thus, XAS is very sensitive to the valence, orbital and spin states of the 3$d$ ions in the initial state. For a thin film on a substrate, XAS can be determined by measuring the drain current due to escaping photoelectrons (Fig. 2(a)). It is referred as the total electron yield (TEY) mode, which is a near surface technique with a probing depth of a few of nm. Via probing the fluorescence yield (FY), one can probe the deeply buried interface or the "bulk" contribution up to a few of hundred nanometers at the $L$ edges of 3d ions. Note that the absorption always provides a measurement that is a weighted average over the probing depth of the detecting modes. With polarized x-rays, one can measure the difference of the absorption of light on passing through a material in two different polarizations, *i.e.*, "dichroism". The common practice is to use either circularly or linearly polarized x-rays, which can be tuned in a modern polarized XAS beamline.

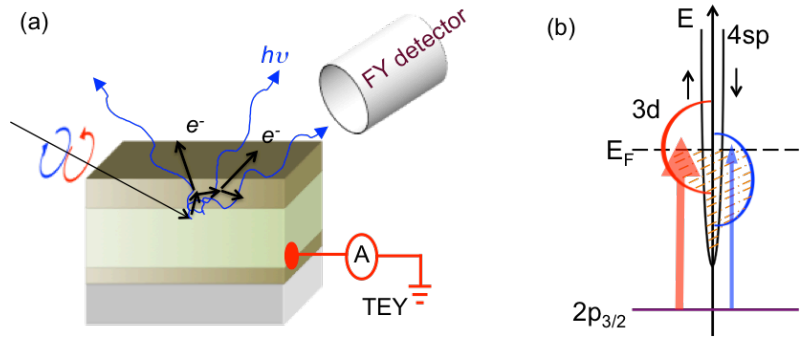

Figure 2. (a) Schematic diagram of a typical configuration of polarized XAS experiments. Circularly (shown in the figure) and linearly polarized x-rays will be used for XMCD and XMLD (XLD) measurements, respectively. Data were normally collected in either the TEY mode (near surface sensitivity) or the FY mode (quasi-bulk sensitivity), or both. (b) X-ray absorption follows the dipole selection rule and spin flips are forbidden in electric dipole transitions. Thus, the spin-split valence shell of a magnetic ion acts as a spin analyzer of the excited photoelectron.

X-Ray Magnetic Circular Dichroism (XMCD) measures the dependence of XAS on the helicity of the circularly polarized x-ray by a magnetic material. It is similar to the Faraday and Kerr effects in the visible light range. Upon absorption, the angular momentum of the circularly polarized x-ray is transferred to the excited photoelectron. If the photoelectron originates from a spin-orbit split core level, *e.g.*, the $p_{3/2}$ level ($L_3$ edge), the angular momentum of the x-ray can be transferred in part to the spin via the spin-orbit coupling. The spin polarization is opposite at the $p_{3/2}$ ($L_3$) and $p_{1/2}$ ($L_2$) levels because they have opposite spin-orbit couplings. Since spin flips are forbidden in electric dipole transitions, the spin-split valence shell of a magnetic ion acts as a spin analyzer of the excited photoelectron (Fig. 2(b)). The transition intensity is approximately proportional to the number of $d$ holes of a given spin subband. The XMCD amplitude scales as $\cos\theta$, where $\theta$ is the angle between the x-ray polarization and the magnetization. Hence, the maximum dichroic effect is observed if the x-ray polarization and the magnetization are parallel and anti-parallel. However, a finite angle (~ 10 deg.) between the beam and the sample plane is typically used to measure in-plane magnetization, because the footprint of the x-ray beam can be much larger than the sample size at low glancing incident angles. A general practice for measuring weak XMCD signal is to flip both the magnetic field and the x-ray helicity. These two are equivalent in principle; therefore it can eliminate the experimental artifacts [22]. However, divergence can arise if there are frozen spin moments that do not flip after reversing the magnetic fields.

The principle of X-Ray Linear Dichroism (XLD) is based on the so-called "search light" effect, where the electric field vector $E$ of linearly polarized x-rays serves as a search light for the number of valence holes along different directions of the atomic volume. The absorption of linearly polarized x-ray depends on the charge anisotropy of the probed ions. This charge anisotropy can arise from an anisotropy in chemical bonding, *i.e.*, by the electrostatic potential, which is referred to natural XLD and can be used to probe the interfacial orbital polarization [17,46]. In the presence of spin order, the spin-orbit coupling leads to preferential charge order relative to the spin direction, which gives rise to the possibility to detect the spin axis. This is called x-ray magnetic linear dichroism (XMLD). Linearly polarized x-rays are only sensitive to axial but not directional properties. Thus, in contrast to XMCD, XMLD has a $\cos^2\theta$ dependence, where $\theta$ is the angle between $E$ and the spin axis. Since linear dichroism can arise from both electric and magnetic asymmetries, one has to distinguish magnetic order effects from ligand field effects, which can be achieved through temperature dependent measurements [30].

There are sum rules that bridge the XAS data with some important microscopic physical quantities. One can determine (a) hole density in the probed valence shell from XAS, (b) spin and orbital moments from XMCD, and (c) anisotropy in the spin-orbit (SO) interaction from XMLD. Below is a brief summary,

- XAS: The integrated intensity of the $L_3$ and $L_2$ resonances is proportional to number of empty $d$ states (holes) for a specific element. Thus, XAS is a great tool to probe the valence state of a particular element, which can be used to study charge transfer [12,33] and electrochemical process [47] across buried interfaces.
- XMCD: The XMCD sum rule connects the experimental integrated intensities of XMCD spectra of the $L_3$ and $L_2$ peaks with the ground-state expectation values of the orbital and spin moments of the absorbing ions [48,49]. This has been widely used to evaluate the interface-induced magnetization [12,33,50]. One has to be cautious when applying the sum rule to determine the spin moment of the early $3d$ transition-metal system due to the mixing of the $L_3$ and $L_2$ peaks [51,52].
- XMLD: The anisotropy in the SO interaction can be obtained by measuring the difference in branching ratio for the $L_{2,3}$ edges for photo polarization parallel and perpendicular to the spin axis [53].

Because of the high x-ray flux and relatively strong light-matter interaction (in comparison to neutrons), synchrotron x-rays allow one to investigate very small quantity of samples. For example, it is straightforward to measure patterned samples, such as tunnel junctions of $100 \times 100\ \mu m^2$ [47,54]. More technical details on magnetism studies using polarized XAS can be found in Ref. [55].

## 3. Materials of Interest

Next we review some recent works on interfacial magnetism from $ABO_3$ perovskite and $A_2BB'O_6$ double perovskite films and heterostructures studied mainly using PNR and polarized X-ray techniques. We focus on interfacial phenomena that are tied to device applications. Topics include: (a) suppressed interfacial magnetization of high spin-polarization oxides adjacent to nonmagnetic insulators, (b) novel exchange coupling between two oxide layers, and (c) emergent interfacial magnetism and its effects in charge transport and magnetization reversal.

### 3.1 High spin-polarization oxides at interfaces

Materials with high spin-polarization at room temperature are of great interest for spin-based advanced sensors and memory applications, for example, as a spin source in magnetic tunnel junctions [56]. The on-going search of high spin-polarization materials has lead to discovery of many interesting compounds, including $CrO_2$, members of the perovskite and double-perovskite families and the Heusler family [57]. Particularly, there are several perovskite and double-perovskite compounds that show half-metallicity with the charge carriers coming mostly from one of the spin subbands, therefore, the charge current is highly spin polarized. Notable examples include $La_{1-x}Sr_xMnO_3$ (LSMO, $x \sim 1/3$) [58,59] and $Sr_2FeMoO_6$ [25,60]. These oxides have been candidates for spintronics devices [61,62]. However, high performance has not been demonstrated at room temperature. In magnetic tunnel junctions composed of $La_{0.67}Sr_{0.33}MnO_3$ and $SrTiO_3$, the tunneling magnetoresistance (TMR) decreases very rapidly when warming up [24]. This has been widely attributed to the suppressed interfacial magnetization [21,22], which can adversely affect the polarization of spin currents through spin-flip scattering or by being a source of the unfavorable minority-spin electrons. There are several important factors that may affect interfacial magnetization of high spin polarization materials, $e.g.$, epitaxial strain [63], cation ordering [23], polar discontinuity [26], octahedral rotation [64] and atomic intermixing [65].

Besides $Sr_2FeMoO_6$, there are several other double pervoskites that are expected to have high spin-polarization with a $Tc$ much higher than room temperatures [25]. $Sr_2CrReO_6$ (SCRO) is particularly interesting because epitaxial SCRO films have a very high $Tc > 500$ K [66] and show semiconducting properties when the Cr/Re cations are highly ordered [67]. In the double-exchange model for double perovskites, there is an antiferromagnetic alignment between the rock-salt ordered Cr and Re spin moments because of the hybridization of the Cr $3d$ and Re $5d$ orbitals via the ligand oxygen, as shown in Fig. 3 (a). Strong spin-orbit coupling is expected due to the presence of Re $5d$ orbitals, therefore the magnetic properties may strongly couple to the lattice changes. Lucy $et\ al.$ [11] performed a combined study of XMCD and high-resolution reciprocal lattice mapping to determine the correlation the strain and the magnetic properties as a function of depth in an 800-nm SCRO film, which was grown on a $SrCr_{0.5}Nb_{0.5}O_3$ (SCNO) buffer layer. The interfacial SCRO region is under 1.1% tensile strain and the film relaxes away from the SCRO/SCNO interface to its bulk lattice parameters. Experiments were conducted at varying glancing-incidence angles. As the incident angle increases, the contribution from the SCRO/SCNO interface increases, which gives rise to the depth sensitivity. A clear correlation between the structural relaxation and the magnetic anisotropy change has been observed. Strain relaxation causes a smooth transition of the magnetic easy axis from out-of-plane near the buffer-layer interface to in-plane far away from the interface.

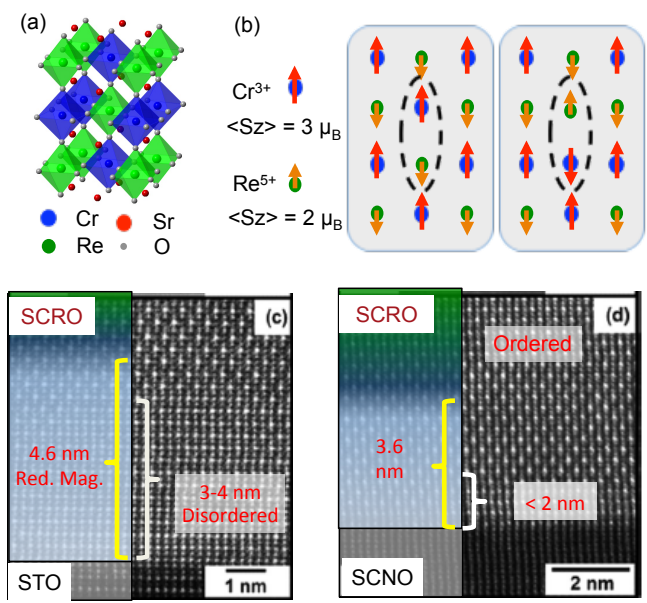

Figure 3. (a) In a fully ordered double perovskite of Sr$_2$CreReO$_6$, the rock-salt ordered Cr and Re spin moments are antiferromagnetic aligned with each other. (b) Based on the double-exchange model for double perovskites, the total magnetization reduces by -2 μ$_B$ for each isolated Cr/Re antisite disorder. PNR experiments found that there are interfacial regions with suppressed magnetization for the SCRO films grown on (c) a bare STO and (d) a SCNO buffer on STO substrate. The suppressed magnetization at interface is correlated with interfacial Cr/Re antisite disorder observed by TEM (adapted from [11]). However, the width of the magnetization suppression is larger than the thickness of the Cr/Re antisite disorder region [23].

Since an 800-nm SCRO film was used in this study, it was too challenging to resolve the magnetic properties with the nanometer depth resolution near the interface via XMCD. Instead, Liu et al. [23] investigated 20-nm SCRO films grown on various substrates/buffers in order to obtain high sensitivity to the interfacial magnetization. X-ray reciprocal space maps have shown that all the 20-nm samples are fully stained to the substrates, therefore the effect from the strain relaxation on the magnetization depth profiles, as found in thicker films, can be excluded. PNR experiments showed that both SCRO films grown on (LaAlO$_3$)$_{0.3}$(Sr$_2$AlTaO$_6$)$_{0.7}$ (LSAT) and SrTiO$_3$ (STO) substrates possess an interfacial layer with reduced magnetization. These two films show comparable widths of magnetization-suppressed regions at the interface, about 4-5 nm. Regions of the Cr/Re antisite disorders at the interfaces had been observed by high-angle annular dark-field scanning transmission electron microscopy in these films [68]. Based on the ferrimagnetic model (Fig. 3(b)) [25], the relative magnetization suppression can be linked to the Cr/Re antisite disorder by, $\frac{m_{disordered}}{m_{ordered}} = 1 - 2x$, where $x$ is the percentage of the Cr/Re disorder. Note that the two films are under different strain states, i.e., 0.8% biaxial compressive strain on LSAT and 0.2% tensile strain on STO, however, the difference in the relative magnetization suppression is small. Therefore, the epitaxial strain seems to only play a weak role on the Cr/Re antisite disorder at interfaces. Interestingly, PNR shows that with a SrCr$_{0.5}$Nb$_{0.5}$O$_3$ (SCNO) buffer layer on STO, the suppressed magnetization region in the SCRO film grown atop decreases to ~ 3.6 nm. Interestingly, as shown in Figs. 3(c) and (d), PNR experiments have revealed that the suppressed magnetization region is ~ 1-2 nm wider than the antisite disorder region for both films gown on STO, regardless whether or not there is a SCNO buffer layer. Therefore, there may exist other mechanisms that suppress the interfacial magnetization of SCRO films. Possibilities include the frustrated exchange coupling between the ordered region and the disordered region and oxygen deficiencies at interfaces. Furthermore, the dominant driving force of the Cr/Re disorder at the interface remains unclear. Note that for spintronics applications, the device performance critically depends on the interfacial properties. As a result, the reduced interfacial magnetization at the interface between double perovskites and nonmagnetic insulators remain an issue for utilizing these materials as high spin-polarization sources, and new strategies are still needed to grow fully ordered layers at interfaces.

The disorder inherent to doping by cation substitution on non-magnetic sites of the complex oxides can also strongly affect spin ordering. For example, neutron diffraction experiments have shown that the Néel temperature of digitally synthesized (layer-by-layer grown) (LaMnO$_3$)$_1$/(SrMnO$_3$)$_2$ superlatttice can reach 320 K, which is much higher than that of the compositionally equivalent random La$_{1/3}$Sr$_{2/3}$MnO$_3$ alloy, which is 250 K [69]. The antiferromagnetic order is A-type, with ferromagnetic ordered Mn sheets in the film plane. Synchrotron X-ray diffraction experiments show that the A-site cation disorder gives rise to an in-plane structural modulation in each ferromagnetic sheet. With the ordering of the A-site cations, this structural modulation is mitigated and driven to long wavelength, which enhances the charge itinerancy within each ferromagnetic sheet and gives rise to a higher spin ordering temperature.

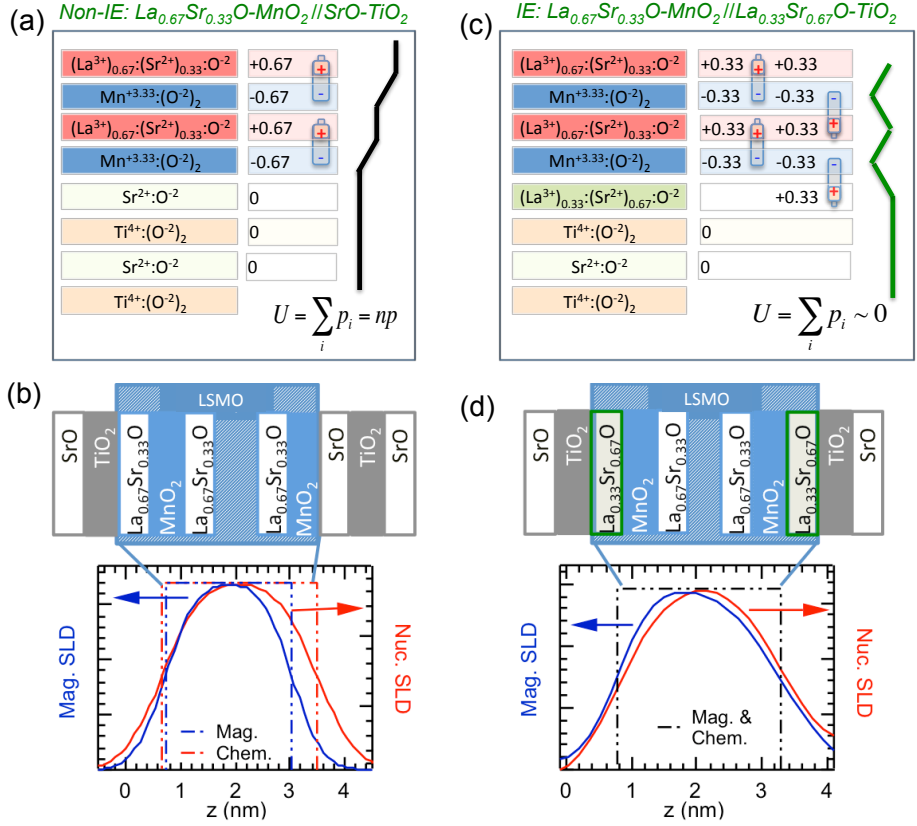

Figure 4. (a) At the ideal $La_{0.67}Sr_{0.33}O$-$MnO_2$-$SrO$-$TiO_2$ interface between LSMO and STO, the polar discontinuity will give rise to the so-called the polarization catastrophe, *i.e.*, the electrostatic potential will build up when stacked the charged $La_{0.67}Sr_{0.33}O$-$MnO_2$ and LaSrO layers sequentially. This may give rise to inferior interfacial properties, which is reduced conductivity and magnetization in this case. One can avoid the polarization catastrophe by interface engineering, for example, replacing the interfacial SrO layer with a $La_{0.33}Sr_{0.67}O$ layer. The depth profiles of the magnetic and nuclear (chemical) scattering length densities for the (b) interface- and (d) non-interface-engineered (IE and Non-IE) $(La,Sr)MnO_3/SrTiO_3$ superlattices as a function of depth (z). Combined Neutron and X-ray scattering measurements confirmed enhanced magnetization at engineered interfaces between $(La,Sr)MnO_3$ and $SrTiO_3$. Figures (b) and (d) are adapted from Ref. [26].

Degradation of magnetic properties at interfaces may also arise from chemical reconstruction driven by a polar discontinuity. When charged planes are stacked, the electrostatic potential may build up (as sketched in Fig. 4 (a)), which will lead to chemical reconstruction at heterointerfaces in order to prevent the polarization catastrophe [9,10]. This happens at the $La_{0.67}Sr_{0.33}MnO_3$ (LSMO) and $SrTiO_3$ (STO) interface [9,26]. LSMO is an double-exchange ferromagnet with a Curie temperature about 360 K, displaying a very high spin polarization at low temperatures [24,59]. The atomic planes of (001) STO are non-polar, while the nominal sheet charge densities of $La_{0.67}Sr_{0.33}O$ plane and the $MnO_2$ plane are $+2/3e$ and $-2/3e$, respectively. Therefore, there is a polar discontinuity at the interface when LSMO being grown on (001) STO. Figure 4 (a) shows the case for one of the two possible interfacial terminations. Using PNR, Huijben *et al.* [26] have determined the magnetization depth profile of a superlattice consisting of five repeating units of 8 unit cells of LSMO and 5 unit cells of STO. Superlattices were used in order to improve the sensitivity to the average interfacial magnetization, due to the enhanced signals around the Bragg peaks. As shown in Fig. 4(b), it was found that the magnetization was confined to a region ~ 6 Å thinner than the chemical thickness of the LSMO layer, which is likely a consequence of the polar discontinuity. Interestingly, the polar discontinuity at the LSMO/STO interface can be removed via interface engineering (IE), for example, by the replacing a SrO layer with of a single $La_{0.33}Sr_{0.67}O$ layer at the $La_{0.67}Sr_{0.33}O$-$MnO_2$-$SrO$-$TiO_2$ interface, as shown in Fig. 4(c). Remarkably, for the case of the IE-superlattice, the magnetic and chemical profiles were commensurate, i.e., the high Mn magnetization persists to the LSMO/STO interface (Fig. 4(d)). Overall, these results show that La-rich interfaces can improve the magnetic properties, which agree with earlier results [21,70]. Therefore, compositionally graded interface can mitigate potentially deleterious electrostatic effects, thus providing a knob to engineer the interfacial properties for layered-oxide based devices.

*3.2 Novel magnetic coupling*

When two oxide materials meet each other forming a heterostructure, charge, spin and orbital degrees of freedoms of *d*-electrons of the components reconstruct in conjunction with the lattice change [5,34]. For example, ferromagnetic features have been observed at interfaces between two different anti-ferromagnetic oxides [19,31,71]. Novel exchange coupling phenomena can also arise at oxide interfaces [18,27,72,73].

One of the prototypical examples of interfacial coupling is exchange-bias phenomena in heterostructures composed of both ferromagnetic (FM) and antiferromagnetic (AFM) components that was initially discovered in 1950s [74]. The characteristic feature of exchange-bias is a shift of the hysteresis loop of the FM layer away from the origin along the field axis, which stems from the unidirectional interfacial coupling between the FM layer and the frozen net moments at the interface that pins the FM layer during the magnetization reversal. Since AFM is typically inert to weak magnetic fields, the uncompensated spin in AFM at the FM/AFM interface can be the source for frozen moments. For magnetic-field sensors and memories based on the spin-valve concept, the exchange bias effect provides an effective mechanism to fix the magnetization of a FM reference layer. Recently exchange-bias has been observed in various other interfaces, such as FM/ferrimagnet [75], FM/FM [27,76] or even AFM/paramagnet [19,72]. The exchange field can be either negative (the exchange field is opposite to the cooling field) or positive (the exchange field has the same sign as the cooling field). Novel exchange bias phenomena are manifested as a shift of a single hysteresis loop or a double-hysteresis loop. Overall, while the exchange-bias effect has been successfully utilized in commercial electronics, there lacks a unified mechanism to explain their origins, which can be system-specific [77]. In this section, we focus on the interfacial exchange coupling in two systems, $SrRuO_3/La_{0.67}Sr_{0.33}MnO_3$ and $La_{0.67}Sr_{0.33}CoO_3/La_{0.67}Sr_{0.33}MnO_3$, both of which consists of a hard FM and a soft FM.

$La_{0.67}Sr_{0.33}MnO_3$ (LSMO) and $SrRuO_3$ (SRO) are both ferromagnetic with $T_c$ about 350 K and 150 K, respectively. LMSO is a soft magnet with a low coercive field (~ 30 Oe), while SRO has a strong uniaxial magnetic anisotropy with a high coercive field of ~ $10^4$ Oe [78]. For a SRO (top)/LSMO (bottom) bilayer grown on a (001) oriented $SrTiO_3$ substrate, Ke *et al.* [76] found that the magnetization of the LSMO layer is positively biased by the SRO layer. With an in-plane magnetic field, the minor hysteresis loop shows that the magnetization reversal of the LSMO layer was shifted along the SRO magnetization direction, which was initially polarized in a large field. This exchange bias effect can be blocked by inserting 2-nm $SrTiO_3$ in between LSMO and SRO layer, which indicates that an antiferromagnetic interfacial coupling rather than the magnetostatic coupling plays the key role [79]. Density functional theory calculations have shown that such an coupling is mediated via the interfacial orbital hybridization of Mn (3*d*) - O (2*p*) - Ru (4*d*) [79,80].

More interestingly, by cooling samples through the Curie temperature of SRO in various magnetic fields, several interesting phenomena have been observed. They are: (i) a single, negatively biased hysteresis loop with small cooling fields; (ii) a single, positively biased hysteresis loop with large cooling fields; (iii) an unusual double-hysteresis loop with intermediate cooling fields [27], as shown in Fig. 5(a) . Ke *et al.* [27] conducted PNR experiments to determine the interfacial spin structures in order to understand the mechanism of this peculiar dependence. The experiments were performed with a two-dimensional position-sensitive detector, which recorded both specular and off-specular reflectivity simultaneously. As illustrated in Fig. 5(b), they found that after cooling in low cooling fields, LSMO magnetization was parallel to the cooling field direction while the SRO magnetization was aligned antiparallel to the cooling field direction. This leads to a negative exchange field on the LSMO layer during the minor hysteresis loop. However, in sufficient large cooling fields, both the SRO magnetization and the LSMO magnetization are parallel to the cooling field direction, giving rise to a positive exchange field. Only the specular signal was clearly observed in these two cases. The case for the intermediate cooling fields, *i.e.*, 500 Oe, was more interesting. There appeared significant off-specular reflectivity that suggested that lateral 180º magnetic domains were formed within the SRO layer. By comparing the experimental off-specular data (Fig. 5(c)) and the modeling (Fig. 5 (d)), Ke *et al.* [27] found that the magnetic domains within SRO layer were of stripes of 200 nm wide and 150-400 μm long. Each SRO stripe domain biases the LSMO layer above, resulting in the double-hysteresis loop feature. The magnetic domain pattern in SRO arose from the detailed balance among the Zeeman energy, the interfacial exchange coupling and the strong uniaxial magnetic anisotropy of the SRO layer. Thus, by changing the cooling field one can manipulate the domain patterns of the SRO layer and accordingly the exchange-bias field applied on the LSMO layer.

Solignac *et al.* [81] performed PNR studies on a similar LSMO (top) /SRO (bottom) bilayer grown by PLD. They found an asymmetric hysteresis loop under a large cooling field and attributed this feature to the presence of two types of domains with different interfacial coupling strength between LSMO and SRO. The two exchange couplings may arise from the coexistence of two types of termination at the LSMO/SRO interfaces, *i.e.*, $RuO_2$-SrO-MnO-$(La_{0.7}Sr_{0.3})O$ and SrO-$RuO_2$-$(La0.7Sr0.3)O$-$MnO_2$. However, this possibility is remote because the high volatility of $RuO_2$ at the high growth temperature and the z-contrast STEM have only found the first termination when growing LSMO on the top of SRO via PLD [82]. Thus, the existence of the two different coupling strengths and its origin need further investigation. At the same time, this feature is not observed in the system studied by Ke *et al.* [76].

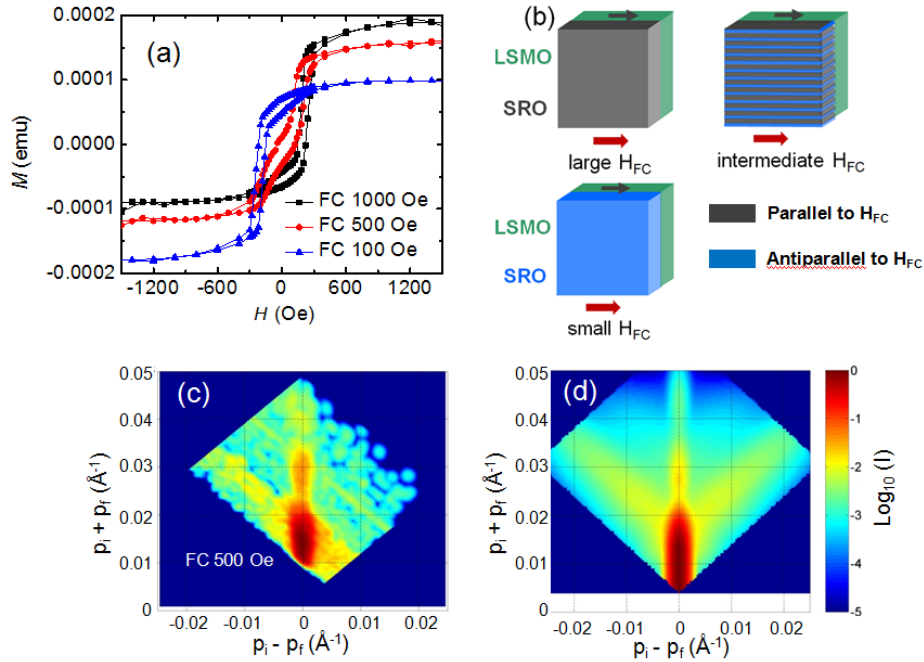

Figure 5: (a) Isothermal hysteresis curves of SRO/LSMO bilayer at $T = 10$ K after field cooling with different magnetic fields. (b) Schematics illustrate the depth profile of the in-plane projection of low temperature spin structure of the bilayer with large, intermediate, and small cooling fields. The experimental (c) and calculated (d) two-dimensional maps of specular and off-specular reflectivity as a function of $Q_z = (p_i-p_f)$ and $(p_i+p_f)$, respectively, with $p_i$ and $p_f$ being the components of the incident and scattered wave vectors perpendicular to the sample surface. Figures are adapted from Ref. [27].

Another interesting exchange-coupling phenomenon was recently reported by Li *et al.* [73] via XMCD measurements in a FM bilayer system composed of $La_{0.7}Sr_{0.3}CoO_3$ (LSCO) with high coercivity and $La_{0.7}Sr_{0.3}MnO_3$ (LSMO) with low coercivity. Thanks to the element sensitivity of XMCD, they found that when the magnetization of the soft LSMO layer reverses, the magnetization of the interfacial LSCO followed it while the bulk part almost remained intact. Such a feature, dubbed as "exchange-spring", is distinct from the exchange-bias behavior previously discussed in LSMO/SRO bilayers where the spins of biasing SRO layer is frozen during the magnetization reversal of the soft LSMO layer. This unexpected result suggests that the exchange coupling between the interfacial LSCO and the LSMO is somehow stronger than that between the interfacial LSCO and the bulk part of the LSCO. The thickness of this interfacial LSCO layer was about 1-2 nm, based on a rough estimation because the TEY mode is a near-surface technique and lacks of a high depth sensitivity. Further investigation using PNR is preferred to underpin the thickness of this interfacial LSCO layer. Note that similar behavior has been reported in an exchange-coupled metallic system consisting of a FM Co layer and an AFM $FeF_2$ layer [83]. The $FeF_2$ layer displays a finite uncompensated magnetization, likely due to defects and/or strains. Using a combined study of PNR and XMCD, Roy *et al.* [83] found the uncompensated $FeF_2$ magnetization at the interface was antiparallel to the Co spins and rotated in conjunction with the Co spins, and the length scale was estimated to be 2 to 3.5 nm.

## 3.3 Emergent interfacial magnetism, and its effects on charge transport and magnetization reversal

In transition-metal perovskites oxide heterostructures, the interfacial *B*-site ions tends to form a strong covalent bond by hybridization via oxygen 2*p* orbital, which may give rise to large magnetic proximity effects (MPE) [18-20,33,84,85]. Because of their interfacial and/or element sensitivity, PNR and XMCD have been frequently used in studying low-dimensional ferromagnetism emerged at oxide interfaces. Chakhailian et al. [33] used XMCD to probe element-specific magnetization in $YBa_2Cu_3O_{7-\delta}$/$La_{0.7}Ca_{0.3}MnO_3$ (YBCO/LCMO) heterostructures, and found emergent Cu net moments at the interface, which are antiferromagnetically coupled to the Mn magnetization, as shown in Fig. 6(a). Later, it was found that the Cu/Mn antiferromagnetic exchange coupling arises from charge transfer and orbital reconstruction at the interfaces based on XLD studies and theoretical calculations [17]. Note that in this scenario, a Mn-O-Cu superexchange path requires the interfacial termination structure of Y-CuO$_2$-BaO-MnO$_2$-(La,Ca)O. Such an interface-induced magnetization have been widely observed at layered-oxide interfaces consisting of individual components with different ground states [18-20], as illustrated in Fig. 6.

YBCO/LCMO heterostructures have been long applied as a model system to study the phase competition between ferromagnetism and superconductivity [28,86-88]. YBCO is a *d*-wave superconductor and LCMO is a half metallic ferromagnet. Earlier work showed that with increasing YCBO layer thickness the saturation magnetization per LCMO layer decreases. Based on PNR experiments, Hoffmann et al. [14] concluded that the reduction of magnetization was related to the magnetization suppression near the YBCO/LCMO interface, which has been attributed to the charge transfer between the two materials and the consequent change in the Mn valence. In the meantime, XMCD work showed that there is also an induced interfacial Cu magnetization that is antiparallel to the Mn magnetization in the adjacent layer [33], which can also contribute to the total magnetization reduction [89]. Thus, suppressed Mn magnetization and induced Cu magnetization coexist at the cuprate-mangantite interfaces, which has been further confirmed by a combined PNR and XMCD study by

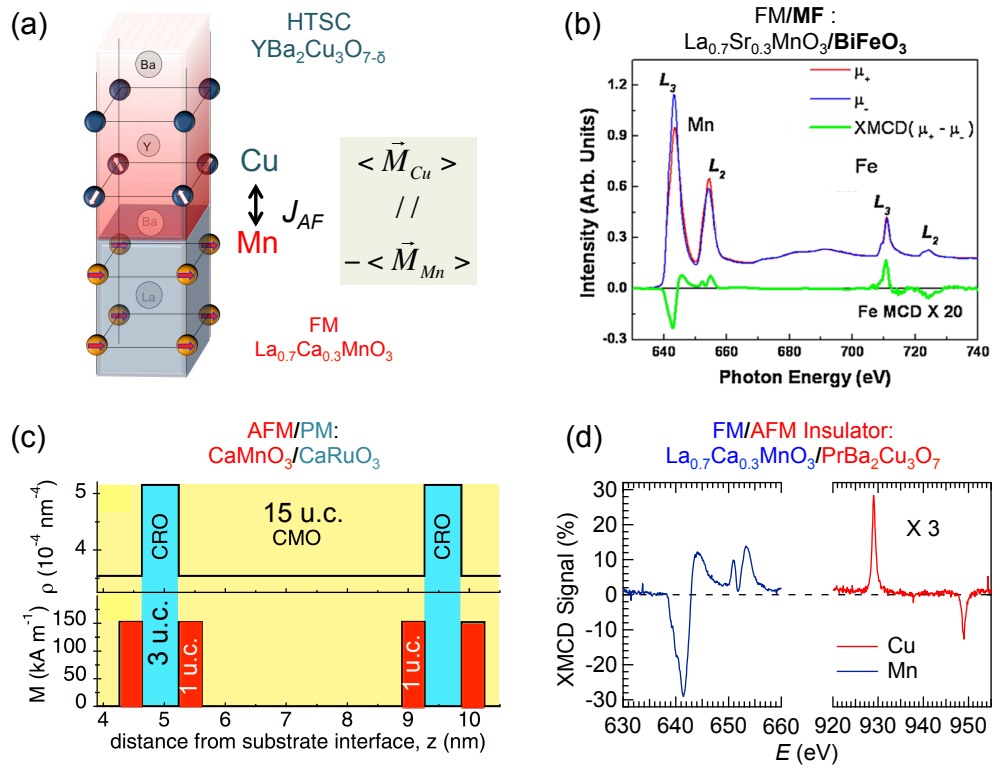

Figure 6: Selected systems that interface induced magnetization have been observed at interfaces between two oxides with different ground states, probed with either polarized neutrons or polarized x-rays. Examples include (a) LCMO (ferromagnet)/YBCO (superconductor) [33], (b) LSMO/BFO (multiferroic) [18], (c) CMO (antiferromagnet)/CRO (paramagnet) [19], and (d) LCMO (ferromagnet)/PBCO (antiferromagnet) [20]. Figures (a-c) are adapted from Refs. [18,19,33].

Satapahty *et al.* [90]. If the net Cu moment was a consequence of the antiferromagnetic superexchange interactions across the interface as generally accepted, there should exist a net Mn moment in the very first $MnO_2$ plane. Therefore, the spin moments of the interfacial LSMO layer are still ordered, at least partially ordered. As mentioned above, the ground state of LCMO is subject to small perturbation, which has been the foundation of a strong magnetoelelctic coupling in hybrid oxide multiferroics [54,91]. Thus one possibility is that it forms an ultrathin A-type AFM manganite interfacial layer at the LCMO/YBCO interface. Although it is technically challenging to directly probe an ultrathin AFM layer of ~ 1-2 u.c. thick, it is achievable by performing XLD experiments at high magnetic fields [92].

Furthermore, it was found that the magnetic proximity effect strongly depends on the electronic state of the manganite layers. By using an insulating ferromagnetic $LaMnO_{3+\delta}$, both the suppressed Mn magnetization and the induced Cu magnetization become largely reduced [90].

On the other hand, MPE effects are still strong when Y in YBCO is partially or fully substituted by Pr, although Pr doping leads to the suppression of conductivity and superconductivity in YBCO [20,93]. Moreover, in the case of $PrBa_2Cu_3O_7$ (PBCO)/LCMO interface, PNR shows that the magnetization of LCMO persists to the interface [20], and the Mn valence state at the interfaces remain close to +3.3 [94], which is in contrast with the case of YBCO/LCMO interfaces [14,89].

Recently, Liu *et al.* [20] have found the induced interfacial Cu magnetization significantly changes the spin dependent transport in the magnetic tunnel junctions (MTJ) consisting of LCMO/PBCO/LCMO trilayers. Contrary to the typically observed a steady increase of the tunnel magnetoresistance with decreasing temperature [24], MTJs of LCMO/PBCO/LCMO trilayers exhibit an anomalous decrease at low temperatures (see Fig. 7 (b)). PNR and XMCD studies on this system show that the saturation magnetization of the LCMO layer increase as the temperature decreases, which rules out the degradation of the ferromagnetic contacts as the cause. Interestingly, the induced net Cu moments at the interface suggests that the spin degeneracy of the conduction band of the PBCO barrier is lifted and thus the barrier becomes spin selective. Based on calculations using the Wentzel-Kramers-Brillouin approximation, Liu et al. [20] showed that the complex temperature dependence can be attributed to the competition between the high positive spin polarization of the manganite electrodes and a negative spin-filter effect from the interfacial Cu magnetization. Note that the interfacial induced magnetization has been widely observed in many transition-metal oxide heterostructures, thus such an interfacial spin-filter effect can be a general feature in layered-oxide based MTJs. Later, Bruno et al. [29] has also identified this effect in MTJs consisting of $La_{0.66}Sr_{0.33}MnO_3$ (LSMO) and $LaFeO_3$. Very interestingly, an induced Ti magnetization has been also observed at $LSMO/SrTiO_3$ interfaces [84]. As mentioned above, magnetic tunnel junctions consisting of LSMO contact and STO barriers have been long used as a model system of oxide spintronics [21,24], therefore a detailed study of the role of Ti net moment on the spin-dependent transport in this system may shed more light on how to enhance the performance at room temperature in this system.

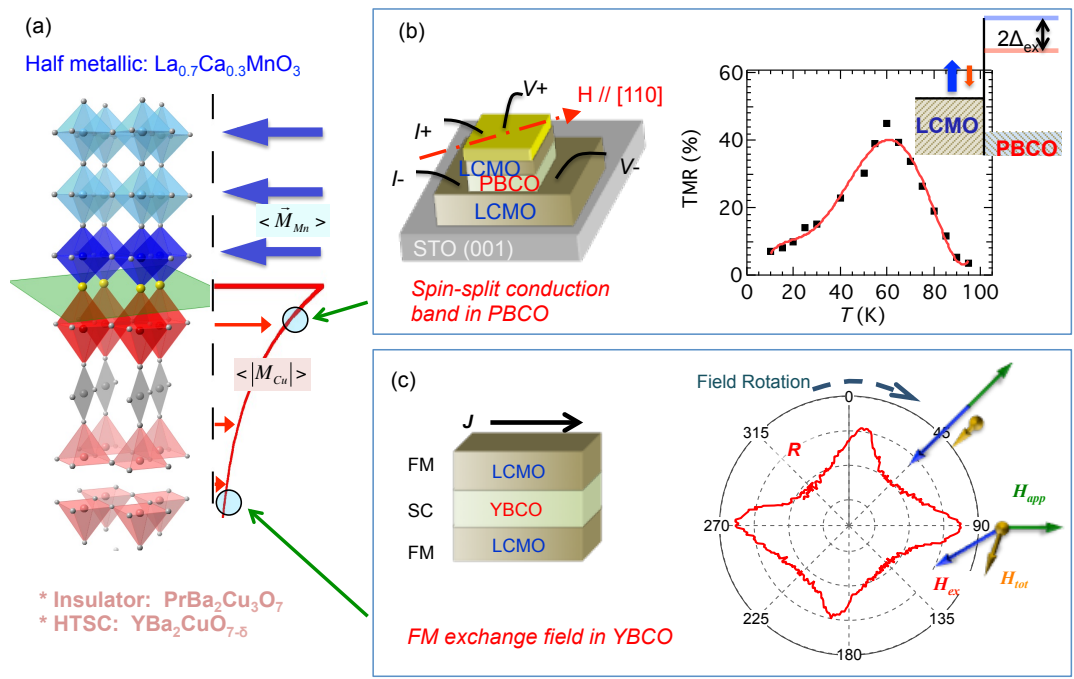

Figure 7: (a) Interface induced magnetization at interfaces between the half metallic manganite and cuprates affects the charge transport in oxide heterostructures. The cuprates can be either an insulator PBCO or a high temperature superconductor YBCO. The Cu magnetization is of interfacial origin and decays exponentially from the interface. This interfacial Cu magnetization can give rise to (b) the interfacial spin-filter effect in magnetic tunnel junctiosn consisting of LCMO/PBCO/LCMO, as well as (c) the interfacial Jaccarino-Peter effect in LCMO/YBCO heterostructures. It is worth noting that in the case of magnetic tunnel junction of LCMO/PBCO/LCMO, the anomalous temperature dependence of TMR is largely due to the induced magnetization in the very first interfacial PBCO layer, where the spin split in conduction band is largest, thus the spin-filter effect is strongest [20]. While in the case of LCMO/YBCO, the observed symmetry, in the angular dependence of the magentoresistance, is due to the tail of the induced ferromagnetic exchange field that competes with the superconductivity in of YBCO. This is because the conductance is dominated by the charge transport in YBCO far away from the interface in the superconducting transition region ($T \sim T_c$), and an external field tips this competition [28]. Figures (b) and (c) are adapted from Refs.[20] and [28], respectively.

The induced Cu interfacial magnetization can also give rise to an unconventional magnetoresistance mechanism in superconducting/ferromagnetic hybrids of YBCO/LCMO. Theory predicts that the induced Cu magnetization decays exponentially from the interface but extends in to the inner YBCO over a length scale of a few of unit cells [95]. The induced Cu net moments means a spin polarization of Cu, which is a consequence of an effective ferromagnetic exchange field on the YBCO. Liu et al. [28] studied superconducting spin-switch structures consisting of trilayers of 15-nm LCMO/ 8-nm YBCO/15-nm LCMO. The two LCMO layers have different anisotropy, likely due to different strain states. The MR in the superconducting transition region was measured during an in-plane field rotation with the magnitude of the field between the anisotropy fields of the two LCMO layers. The rotational magnetoresistance shows pronounced features, such as a quasi-four-fold symmetry and an angular hysteresis between clockwise and anti-clockwise rotations (see Fig. 7 (c)). In contradiction to several proposed scenarios, there is no clear correlation between the MR and the magnetization alignment between the two LCMO layers. Via PNR, Liu et al. [28] found that during the field rotation, the magnetization of the bottom LCMO closely tracked the field direction, but the top LCMO layer remained almost intact. By comparing the symmetries of the magnetization structure and magnetoresistance during the field rotation, the observed magnetoresistance can be explained by the total field in the central YBCO region, which results from the superposition of the applied magnetic field and the exponential tail of the aforementioned exchange-field. This observation is reminiscent of the Jaccarino-Peter effect of the magnetic-field induced superconductivity [96]. Soon after this study, this interfacial Jaccarino-Peter effect has been also observed in non-oxide based devices [97]. It is worth noting that the induced Cu magnetization has also been found at the interface between $La_{0.66}Sr_{0.33}MnO_3$ and $La_{1.85}Sr_{0.15}CuO_4$ (LSCO) recently [85]. It will be of interest to investigate the effect of the induced Cu magnetization on the superconductivity in LSCO.

Recent work also show that the induced interfacial magnetism can affect the magnetization reversal process [18,19,72]. For instance, He et al. [19] studied $CaRuO_3/CaMnO_3$ (CRO/CMO) superlattices, where $CaRuO_3$ is a paramagnetic metal while $CaMnO_3$ is an antiferromagnetic insulator. From mangetometery studies, they observed finite magnetization and the exchange bias effect. They conducted PNR experiments and unraveled ferromagnetism induced within the first unit cell of $CaMnO_3$ at the interface. However, x-ray resonant magnetic scattering experiments performed by Freeland et al. [98] on a similar $CaRuO_3/CaMnO_3$ heterostructure showed that the canted Mn spins penetrates 3–4 unit cells into $CaMnO_3$ from the interface. The discrepancy in the length scale is unclear yet, which may be related to subtle difference of the interfacial structures of the samples grown by different groups. Note that the films from both groups were grown by PLD but on different substrates. Samples used by Freeland et al. were grown on $LaAlO_3$ and the films were completely lattice matched to the substrate. On the other hand, He et al. used $SrTiO_3$ substrates thus the lattice mismatch between the film and the substrate is much larger, and consequently the CMO layers were structurally relaxed. However, assuming that the interface Mn magnetization is from the double exchange enabled by electrons leaking from CRO to CMO, the density-functional theory calculations predict a length scale of one unit cell [99], agreeing with the PNR results from He et al. [19]. Regardless the length scale, the hysteresis loop of such an interface induced ferromagnetic layer can be exchange coupled with the rest antiferromagnetic $CaMnO_3$ layers, which explains the observed exchange bias behavior [19].

Similarly, Gibert et al. [72] have observed an exchange-bias behavior in $(LaMnO_3)_n/(LaNiO_3)_m$ (LNO/LMO) superlattices grown on (111) oriented $SrTiO_3$ substrates. The result appears quite unexpected since stoichiometric bulk LMO is an A-type antiferromagnet ($T_N \sim 140$ K) and bulk LNO is a paramagnetic metal. Nevertheless, it has been found that ultrathin $LaNiO_3$ films exhibit magnetotransport properties that are tied to either a spin-class state or an insulating antiferromagnetic state [100]. And LMO thin films exhibits ferromagnetism with a Curie temperature of ∼ 200 K [12,72], possibly due to oxygen vacancies and/or strain. Later, Hoffman *et al.* [12] performed XAS and XMCD studies on (001) LNO/LMO superlattices and showed clear evidence of the interfacial charge transfer between $LaMnO_3$ and $LaNiO_3$ and a large net Ni moment. Based on these facts, the exchange bias effect in LNO/LMO superlattices can be understood in the conventional picture of a FM/AFM interface. According to the first-principle density functional theory calculations, sizeable Ni magnetization is expected provided Mn-to-Ni charge transfer across the interface [101]. However, Dong and Dagotto [102] have employed a hybrid two-orbital tight-binding model and shown that the quantum confinement effect play a more dominated role on the Ni ferromagnetism in LNO in the (111)-stacking superlattices, but this model does not predict a sizable Ni magnetization the (001) stacking superlattices.

Very interestingly, in multiferroic heterostructures consisting of $BiFeO_3$ (BFO) and $La_{0.7}Sr_{0.3}MnO_3$, interface-induced large net BFO magnetization has been observed by both XMCD [18] and PNR [32] experiments. The development of the BFO net moment is strongly associated with the onset of an exchange bias [18], the sign of which appears reversibly switchable upon ferroelectric poling of the BFO [103]. So far, the effects are only observed at low temperatures. These results are relevant to device applications because it promises a strong interfacial magnetoelectric effect, which can potentially enable the electrical-field control of magnetism at room temperature in the near future.

## 4. Summary and Perspective

Neutron and x-ray studies on interfacial magnetism in layered oxide heterostructures have been dramatically expanded in the last a few of years. We have reviewed several examples to highlight some capabilities of polarized neutrons and x-rays and the emerging research themes. We put the emphasis on systems related to high spin polarization oxides, such as hole-doped manganites and high Curie temperature double perovskites. Multiferroic heterostructures are of great application value [35,104] but have been only briefly discussed above. Several groups have conducted in-situ neutron and x-ray experiments with applying voltage biases to investigate the interfacial magnetoelectric coupling in FE/FM oxide heterostructures, e.g., Ref. [54,105], and more is undergoing. The research is driven by the search of energy-efficient spintronics without using an external magnetic field. At the same time, such hybrids may extend or even replace current CMOS technology since one can control materials properties beside the conductance. In this later theme, electrostatic gating of ultrathin oxide films has been extensively studied to create thermally inaccessible phases and change the magnetic and transport properties [106,107]. This approach is scientifically beautiful since it does not introduce chemical disorders. On the other hand, one can use an oxygen ion conductor as gate dielectrics to gain an extra knob on oxygen vacancy concentration, and then one can achieve memristive behavior, which is of high values to applications.

It is worth noting that at the nanoscale, the boundary between materials physics and electrochemistry is very blurring [108]. A notable example is the electrolyte gating of $VO_2$ with an ionic liquid [109], where oxygen vacancies in $VO_2$ can be created by applying an electric field thus the metal-insulator transition temperature can be shifted accordingly. As mentioned previously, in many families of layered transition metal oxides, *e.g.*, superconducting cuprate [110] and magnetoresistive manganite [1], the physical properties are very sensitive to the oxygen nonstoichiometry. Thus by controlling the oxygen vacancies, one is able to manipulate the physical properties, such as magnetization [111]. This can be achieved by putting an ultrathin oxide films of interest in adjacent to an oxygen conductor, and then one can drive oxygen ions into and out from the oxide films via applying an electric field, as illustrated in Fig. 8(a). This redox process across interfaces can be realized with solid gating, which is compatible with current semiconductor technology. This strategy has been recently employed in a metallic system consisting of ultrathin ferromagnetic Co films and the oxygen conductor $Gd_2O_{3-x}$ [47]. XAS and XMCD have shed light on the redox process at the interfaces. Note that a high incident angle and the FY model were used during the experiments in order to enhance the sensitivity of the deeply buried Co layer, as shown in Fig. 8(b). Combined with magnetotransport studies, a reversible control of Co magnetism has been demonstrated. The observed effect is very pronounced in terms of changes of the magnetic anisotropy. The ultrathin Co layer can be switched between the pure metallic Co state and the $Co^{2+}$ state (in form of CoO) with a low bias of 5 V. Thus the magnetization, the magnetic anisotropy and the charge, optical, and thermal conductivities can all be largely manipulated with small electrical fields. Certainly, it is of high interest to extend this strategy to complex oxides, and polarized neutrons and x-rays will be among the most effective probes to investigate the redox process.

As mentioned earlier, the competition between magnetization and superconductivity in oxide heterostructures is worth further investigation. In particular, spin spiral structures can be achieved and manipulated via exchange coupling in FM/AFM bilayers, or soft FM/hard FM bilayers. This structure has been used to explore the competition between noncollinear magnetization and conventional superconductivity [112]. Similar approaches can be used to search and study exotic quantum states in heterostructures of half-metallic oxides and high temperature superconductor. In this case, PNR with polarization analysis can contribute to determine the depth profile of spin spiral structures [42] in order to build up the correlation between the magnetic structure and the superconductivity at the interface.

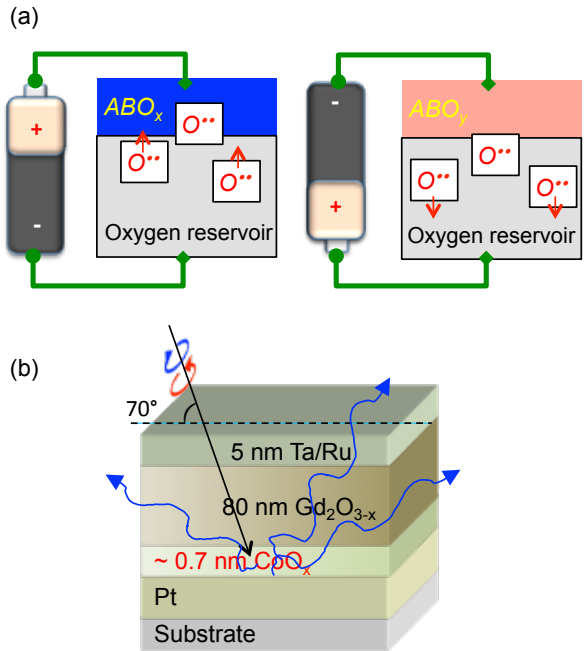

Figure 8: (a) By putting an ultrathin oxide films of interest adjacent to an oxygen conductor, the oxygen ion conductor will behave as an oxygen reservoir and oxygen ions can be driven into and out from the ultrathin oxide films via applying an electric field. (b) XAS and XMCD (FY mode) have been used to investigate an ultrathin Co film in adjacent oxygen ion conductor $Gd_2O_{3-x}$. A large, reversible, and nonvolatile change in magnetism has been found tie to the redox of Co via electric fields [47].

Antiferromagnetic spintronics is a new and rapidly developing field, but there is very limited work focusing on antiferromagnetism in layered oxide films and heterostructures. One notable work is the enhanced Néel temperatures in cation ordered manganite films studied via neutron diffraction [69]. On the other hand, XMLD is very powerful to determine the spin axis of antiferromagnetism and to search for emerging interfacial antiferromagnetism, e.g., between two nonmagnetic oxides. In principle, antiferromagnetism reflectometry can be developed based on resonant magnetic scattering to determine the depth profile of the spin axis of an AFM.

We conclude by summarizing that interface-induced magnetism have been widely and unexpectedly observed in many oxide heterostructures with various ground states (See Fig. 6), like half-metal and high-temperature superconductor [33,85,113], ferromagnetic metal and antiferromagnetic insulator [20,29,114], antiferromagnetic insulator and paramagnetic metal [19,98], ferromagnetic metal and nonmagnetic insulator [84], ferromagnet and ferroelectric [115], ferromagnet and mutiferroics [18,32], high-temperature superconductor and even between two nonmagnetic insulators [116,117]. Some systems do not have clear connections with applications but have deepened our understanding in physics of strongly correlated materials, for example $LaAlO_3/SrTiO_3$, in which 2D electron gas, superconducting and/or ferromagnetic behaviors have all been serendipitously observed [116]. Some others are closely tied to applications including magnetic tunnel junctions and artificial multiferroic heterostructures. Such interface-induced magnetism can play important roles in the charge transport and magnetization reversal of the oxide heterostructures. Recent work have also shown that the interface-induced magnetism are tunable via strain engineering in cases of SRO/LSMO [114] and LSMO/STO [84]. Therefore, the interface-induced magnetism in layered oxide heterostructures provides a knob for engineering oxide spintronics, and it is fair to stress that polarized neutron and x-ray studies have been and will continue playing a crucial role in this exciting research area.

## Acknowledgments

We sincerely acknowledge Dr. M. Fitzsimmons (ORNL) for carefully reading our manuscript and giving valuable comments. We gratefully thank many friends and collaborators for discussions in the past years, including but not limited to J.A. Borchers (NCNR), H. Boschker (MPI-FKF), C.B. Eom (UW-Madison), J.W. Freeland (APS), J. Lucy (OSU), M.S. Rzchowski (UW-Madison), J. Santamaria (CUM), P. Shafer (ALS), SGE te Velthuis (ANL), W. Wang (U. Arizona), H. Zhou (APS) and C.H. Zhu (ALS). We also acknowledge Ms. K. Bethea (ORNL) for her assistance on Fig. 4. Liu is supported by the Division of Scientific User Facilities of the Office of Basic Energy Sciences, US Department of Energy. Ke is supported by start-up funds at Michigan State University.

## References

[1]   J. F. Mitchell, D. N. Argyriou, C. D. Potter, D. G.

Hinks, J. D. Jorgensen, and S. D. Bader, Phys. Rev., B Condens. Matter **54**, 6172 (1996).
[2] O. Chmaissem, B. Dabrowski, S. Kolesnik, J. Mais, J. Jorgensen, and S. Short, Phys. Rev. B **67**, 094431 (2003).
[3] M. B. Salamon and M. Jaime, Rev. Mod. Phys. **73**, 583 (2001).
[4] E. O. WOLLAN and W. C. KOEHLER, Phys. Rev. **100**, 545 (1955).
[5] H. Y. Hwang, Y. Iwasa, M. Kawasaki, B. Keimer, N. Nagaosa, and Y. Tokura, Nat Mater **11**, 103 (2012).
[6] A. Bhattacharya and S. J. May, Annu. Rev. Mater. Res. **44**, 65 (2014).
[7] P. Zubko, S. Gariglio, M. Gabay, P. Ghosez, and J.-M. Triscone, Annu. Rev. Condens. Matter Phys. **2**, 141 (2011).
[8] J. Chakhalian, J. W. Freeland, A. J. Millis, C. Panagopoulos, and J. M. Rondinelli, Rev. Mod. Phys. **86**, 1189 (2014).
[9] H. Boschker, J. Verbeeck, R. Egoavil, S. Bals, G. van Tendeloo, M. Huijben, E. P. Houwman, G. Koster, D. H. A. Blank, and G. Rijnders, Adv. Funct. Mater. **22**, 2235 (2012).
[10] N. Nakagawa, H. Y. Hwang, and D. A. Muller, Nat Mater **5**, 204 (2006).
[11] J. M. Lucy, A. J. Hauser, H. L. Wang, J. R. Soliz, M. Dixit, R. E. A. Williams, A. Holcombe, P. Morris, H. L. Fraser, D. W. McComb, P. M. Woodward, and F. Y. Yang, Appl. Phys. Lett. **103**, 042414 (2013).
[12] J. Hoffman, I. C. Tung, B. B. Nelson-Cheeseman, M. Liu, J. W. Freeland, and A. Bhattacharya, Phys. Rev. B **88**, 144411 (2013).
[13] S. Yunoki, A. Moreo, E. Dagotto, and S. Okamoto, Phys. Rev. B **76**, 064532 (2007).
[14] A. Hoffmann, S. G. E. te Velthuis, Z. Sefrioui, J. Santamaría, M. R. Fitzsimmons, S. Park, and M. Varela, Phys. Rev. B **72**, 140407(R) (2005).
[15] S. J. May, J. W. Kim, J. M. Rondinelli, E. Karapetrova, N. A. Spaldin, A. Bhattacharya, and P. J. Ryan, Phys. Rev. B **82**, 014110 (2010).
[16] A. Vailionis, H. Boschker, Z. Liao, J. R. A. Smit, G. Rijnders, M. Huijben, and G. Koster, Appl. Phys. Lett. **105**, 131906 (2014).
[17] J. Chakhalian, J. W. Freeland, H. U. Habermeier, G. Cristiani, G. Khaliullin, M. van Veenendaal, and B. Keimer, Science **318**, 1114 (2007).
[18] P. Yu, J. S. Lee, S. Okamoto, M. D. Rossell, M. Huijben, C. H. Yang, Q. He, J. X. Zhang, S. Y. Yang, M. J. Lee, Q. M. Ramasse, R. Erni, Y. H. Chu, D. A. Arena, C. C. Kao, L. W. Martin, and R. Ramesh, Phys. Rev. Lett. **105**, 027201 (2010).
[19] C. He, A. J. Grutter, M. Gu, N. D. Browning, Y. Takamura, B. J. Kirby, J. A. Borchers, J. W. Kim, M. R. Fitzsimmons, X. Zhai, V. V. Mehta, F. J. Wong, and Y. Suzuki, Phys. Rev. Lett. **109**, 197202 (2012).
[20] Y. Liu, F. A. Cuellar, Z. Sefrioui, J. W. Freeland, M. R. Fitzsimmons, C. Leon, J. Santamaría, and S. G. E. te Velthuis, Phys. Rev. Lett. **111**, 247203 (2013).
[21] H. Yamada, Y. Ogawa, Y. Ishii, H. Sato, M. Kawasaki, H. Akoh, and Y. Tokura, Science **305**, 646 (2004).
[22] J. W. Freeland, J. J. Kavich, K. E. Gray, L. Ozyuzer, H. Zheng, J. F. Mitchell, M. P. Warusawithana, P. Ryan, X. Zhai, R. H. Kodama, and J. N. Eckstein, Journal of Physics: Condensed Matter **19**, 315210 (2007).
[23] Y. Liu, J. M. Lucy, A. Glavic, H. Ambaye, V. Lauter, F. Y. Yang, and S. G. E. te Velthuis, Phys. Rev. B **90**, 104416 (2014).
[24] M. Bowen, M. Bibes, A. Barthélémy, J. P. Contour, A. Anane, Y. Lemaître, and A. Fert, Appl. Phys. Lett. **82**, 233 (2003).
[25] D. Serrate, J. M. D. Teresa, and M. R. Ibarra, Journal of Physics: Condensed Matter **19**, 023201 (2006).
[26] M. Huijben, Y. Liu, H. Boschker, and V. Lauter, Adv. Mater. (2015).
[27] X. Ke, L. J. Belenky, V. Lauter, H. Ambaye, C. W. Bark, C. B. Eom, and M. S. Rzchowski, Phys. Rev. Lett. **110**, 237201 (2013).
[28] Y. Liu, C. Visani, N. M. Nemes, M. R. Fitzsimmons, L. Y. Zhu, J. Tornos, M. Garcia-Hernandez, M. Zhernenkov, A. Hoffmann, C. Leon, J. Santamaría, and S. G. E. te Velthuis, Phys. Rev. Lett. **108**, 207205 (2012).
[29] F. Y. Bruno, M. N. Grisolia, C. Visani, S. Valencia, M. Varela, R. Abrudan, J. Tornos, A. Rivera-Calzada, A. A. Ünal, S. J. Pennycook, Z. Sefrioui, C. Leon, J. E. Villegas, J. Santamaría, A. Barthélémy, and M. Bibes, Nature Communications **6**, 6306 (2015).
[30] E. Arenholz, G. van der Laan, F. Yang, N. Kemik, M. D. Biegalski, H. M. Christen, and Y. Takamura, Appl. Phys. Lett. **94**, 072503 (2009).
[31] A. J. Grutter, H. Yang, B. J. Kirby, M. R. Fitzsimmons, J. A. Aguiar, N. D. Browning, C. A. Jenkins, E. Arenholz, V. V. Mehta, U. S. Alaan, and Y. Suzuki, Phys. Rev. Lett. **111**, 087202 (2013).
[32] S. Singh, J. T. Haraldsen, J. Xiong, E. M. Choi, P. Lu, D. Yi, X. D. Wen, J. Liu, H. Wang, Z. Bi, P. Yu, M. R. Fitzsimmons, J. L. MacManus-Driscoll, R. Ramesh, A. V. Balatsky, J.-X. Zhu, and Q. X. Jia, Phys. Rev. Lett. **113**, 047204 (2014).
[33] **2**, 244 (2006).
[34] J. Mannhart and D. G. Schlom, Science **327**, 1607 (2010).
[35] S. Fusil, V. Garcia, A. Barthélémy, and M. Bibes, Annu. Rev. Mater. Res. **44**, 91 (2014).
[36] M. R. Fitzsimmons, in *Modern Techniques for Characterizing Magnetic Materials* (Springer US, New York, 2005), pp. 107–155.
[37] V. Lauter-Pasyuk, H. J. Lauter, B. P. Toperverg, L. Romashev, and V. Ustinov, Phys. Rev. Lett. **89**, 167203 (2002).
[38] Y. Liu, M. Iavarone, A. Belkin, G. Karapetrov, V. Novosad, M. Zhernenkov, Q. Wang, M. R. Fitzsimmons, V. Lauter, and S. G. E. te Velthuis, arXiv arXiv:1410.5520 (2014).
[39] **107**, (2011).
[40] B. J. Kirby, D. Kan, A. Luykx, M. Murakami, D. Kundaliya, and I. Takeuchi, J. Appl. Phys. **105**, 07D917 (2009).

[41] L. G. Parratt, Phys. Rev. **95**, 359 (1954).
[42] Y. Liu, S. Te Velthuis, J. S. Jiang, Y. Choi, and S. D. Bader, Phys. Rev. B **83**, 174418 (2011).
[43] S. Singh, M. R. Fitzsimmons, T. Lookman, and H. Jeen, Phys. Rev. B **85**, 214440 (2012).
[44] M. R. Fitzsimmons, B. J. Kirby, N. W. Hengartner, F. Trouw, M. J. Erickson, S. D. Flexner, T. Kondo, C. Adelmann, C. J. Palmstrøm, P. A. Crowell, W. C. Chen, T. R. Gentile, J. A. Borchers, C. F. Majkrzak, and R. Pynn, Phys. Rev. B **76**, 245301 (2007).
[45] C. F. Majkrzak, K. V. O'Donovan, and N. F. Berk, in *Neutron Scattering From Magnetic Materials* (Elsevier, 2006), pp. 397–471.
[46] N. Nücker, E. Pellegrin, P. Schweiss, J. Fink, S. Molodtsov, C. Simmons, G. Kaindl, W. Frentrup, A. Erb, and G. Müller-Vogt, Phys. Rev., B Condens. Matter **51**, 8529 (1995).
[47] C. Bi, Y. Liu, T. Newhouse-Illige, M. Xu, M. Rosales, J. W. Freeland, O. Mryasov, S. Zhang, S. G. E. te Velthuis, and W. G. Wang, Phys. Rev. Lett. **113**, 267202 (2014).
[48] B. T. THOLE, P. Carra, F. Sette, and G. van der Laan, Phys. Rev. Lett. **68**, 1943 (1992).
[49] P. Carra, B. T. THOLE, M. Altarelli, and X. Wang, Phys. Rev. Lett. **70**, 694 (1993).
[50] J. Garcia-Barriocanal, J. C. Cezar, F. Y. Bruno, P. Thakur, N. B. Brookes, C. Utfeld, A. Rivera-Calzada, S. R. Giblin, J. W. Taylor, J. A. Duffy, S. B. Dugdale, T. Nakamura, K. Kodama, C. Leon, S. Okamoto, and J. Santamaría, Nature Communications **1**, 1 (2010).
[51] C. Piamonteze, P. Miedema, and F. M. F. de Groot, Phys. Rev. B **80**, 184410 (2009).
[52] A. Scherz, H. Wende, C. Sorg, K. Baberschke, J. Minr, D. Benea, and H. Ebert, Physica Scripta **T115**, 586 (2005).
[53] G. van der Laan, Phys. Rev. Lett. **82**, 640 (1999).
[54] C. A. F. Vaz, J. Hoffman, Y. Segal, J. W. Reiner, R. D. Grober, Z. Zhang, C. H. Ahn, and F. J. Walker, Phys. Rev. Lett. **104**, 127202 (2010).
[55] J. STOHR and H. C. Siegmann, *Magnetism*, 1st ed. (Springer Berlin Heidelberg, Berlin Heidelberg New York, 2006).
[56] J. M. D. Coey and C. L. Chien, MRS Bull. **28**, 720 (2003).
[57] C. Felser and G. H. Fecher, *Spintronics: From Materials to Devices* (Springer Netherlands, Dordrecht, 2013).
[58] M. Bowen, A. Barthélémy, M. Bibes, E. Jacquet, J. P. Contour, A. Fert, F. Ciccacci, L. Duò, and R. Bertacco, Phys. Rev. Lett. **95**, 137203 (2005).
[59] J. H. Park, E. Vescovo, H. J. Kim, C. Kwon, R. Ramesh, and T. Venkatesan, Nature **392**, 794 (1998).
[60] K. I. Kobayashi, T. Kimura, H. Sawada, and K. Terakura, Nature **395**, 677 (1998).
[61] M. Bibes and A. Barthélémy, Ieee Transactions on Electron Devices **54**, 1003 (2007).
[62] H. Asano, N. Koduka, K. Imaeda, M. Sugiyama, and M. Matsui, IEEE Trans. Magn. **41**, 2811 (2015).
[63] L. Abad, V. Laukhin, S. Valencia, A. Gaup, W. Gudat, L. Balcells, and B. Martínez, Adv. Funct. Mater. **17**, 3918 (2007).
[64] X. Zhai, L. Cheng, Y. Liu, C. M. Schlepütz, S. Dong, H. Li, X. Zhang, S. Chu, L. Zheng, J. Zhang, A. Zhao, H. Hong, A. Bhattacharya, J. N. Eckstein, and C. Zeng, Nature Communications **5**, 4283 (2014).
[65] L. F. Kourkoutis, J. H. Song, H. Y. Hwang, and D. A. Muller, Proceedings of the National Academy of Sciences **107**, 11682 (2010).
[66] H. Kato, T. Okuda, Y. Okimoto, Y. Tomioka, Y. Takenoya, A. Ohkubo, M. Kawasaki, and Y. Tokura, Appl. Phys. Lett. **81**, 328 (2002).
[67] A. J. Hauser, J. R. Soliz, M. Dixit, R. E. A. Williams, M. A. Susner, B. Peters, L. M. Mier, T. L. Gustafson, M. D. Sumption, H. L. Fraser, P. M. Woodward, and F. Y. Yang, Phys. Rev. B **85**, 161201 (2012).
[68] J. M. Lucy, A. J. Hauser, Y. Liu, H. Zhou, Y. Choi, D. Haskel, S. G. E. te Velthuis, and F. Y. Yang, Phys. Rev. B **91**, 094413 (2015).
[69] S. J. May, P. J. Ryan, J. L. Robertson, J. W. Kim, T. S. Santos, E. Karapetrova, J. L. Zarestky, X. Zhai, S. G. E. te Velthuis, J. N. Eckstein, S. D. Bader, and A. Bhattacharya, Nat Mater **8**, 892 (2009).
[70] J. J. Kavich, M. P. Warusawithana, J. W. Freeland, P. Ryan, X. Zhai, R. H. Kodama, and J. N. Eckstein, Phys. Rev. B **76**, 014410 (2007).
[71] T. S. Santos, B. J. Kirby, S. Kumar, S. J. May, J. A. Borchers, B. B. Maranville, J. Zarestky, S. G. E. te Velthuis, J. van den Brink, and A. Bhattacharya, Phys. Rev. Lett. **107**, 167202 (2011).
[72] M. Gibert, P. Zubko, R. Scherwitzl, J. Iñiguez, and J.-M. Triscone, Nat Mater **11**, 195 (2012).
[73] B. Li, R. V. Chopdekar, E. Arenholz, A. Mehta, and Y. Takamura, Appl. Phys. Lett. **105**, 202401 (2014).
[74] W. H. Meiklejohn and C. P. Bean, Phys. Rev. **102**, 1413 (1956).
[75] S. Mangin, G. Marchal, and B. Barbara, Phys. Rev. Lett. **82**, 4336 (1999).
[76] X. Ke, M. S. Rzchowski, L. J. Belenky, and C. B. Eom, Appl. Phys. Lett. **84**, 5458 (2004).
[77] M. R. Fitzsimmons and I. K. Schuller, Journal of Magnetism and Magnetic Materials **350**, 199 (2014).
[78] Q. Gan, R. A. Rao, C. B. Eom, L. Wu, and F. Tsui, J. Appl. Phys. **85**, 5297 (1999).
[79] M. Ziese, I. Vrejoiu, E. Pippel, P. Esquinazi, D. Hesse, C. Etz, J. Henk, A. Ernst, I. V. Maznichenko, W. Hergert, and I. Mertig, Phys. Rev. Lett. **104**, 167203 (2010).
[80] Y. Lee, B. Caes, and B. N. Harmon, Journal of Alloys and Compounds **450**, 1 (2008).
[81] A. Solignac, R. Guerrero, P. Gogol, T. Maroutian, F. Ott, L. Largeau, P. Lecoeur, and M. Pannetier-Lecoeur, Phys. Rev. Lett. **109**, 027201 (2012).
[82] G. Rijnders, D. H. A. Blank, J. Choi, and C.-B. Eom, Appl. Phys. Lett. **84**, 505 (2004).
[83] S. Roy, M. R. Fitzsimmons, S. Park, M. Dorn, O. Petracic, I. V. Roshchin, Z.-P. Li, X. Batlle, R. Morales, A. Misra, X. Zhang, K. Chesnel, J. B. Kortright, S. K. Sinha, and I. K. Schuller, Phys. Rev. Lett. **95**, 047201 (2005).

[84] **106**, 147205 (2011).
[85] G. M. De Luca, G. Ghiringhelli, C. A. Perroni, V. Cataudella, F. Chiarella, C. Cantoni, A. R. Lupini, N. B. Brookes, M. Huijben, G. Koster, G. Rijnders, and M. Salluzzo, Nature Communications **5**, 5626 (2014).
[86] J. W. Freeland, J. Chakhalian, H. U. Habermeier, G. Cristiani, and B. Keimer, Appl. Phys. Lett. **90**, 242502 (2007).
[87] V. Peña, Z. Sefrioui, D. Arias, C. Leon, J. Santamaría, J. Martinez, S. te Velthuis, and A. Hoffmann, Phys. Rev. Lett. **94**, 057002 (2005).
[88] M. Kasai, Y. Kanke, T. Ohno, and Y. Kozono, J. Appl. Phys. **72**, 5344 (1992).
[89] J. Stahn, J. Chakhalian, C. Niedermayer, J. Hoppler, T. Gutberlet, J. Voigt, F. Treubel, H. U. Habermeier, G. Cristiani, B. Keimer, and C. Bernhard, Phys. Rev. B **71**, 140509 (2005).
[90] D. K. Satapathy, M. A. Uribe-Laverde, I. Marozau, V. K. Malik, S. Das, T. Wagner, C. Marcelot, J. Stahn, S. Brück, A. Rühm, S. Macke, T. Tietze, E. Goering, A. Frano, J. H. Kim, M. Wu, E. Benckiser, B. Keimer, A. Devishvili, B. P. Toperverg, M. Merz, P. Nagel, S. Schuppler, and C. Bernhard, Phys. Rev. Lett. **108**, 197201 (2012).
[91] J. D. Burton and E. Y. Tsymbal, Phys. Rev. B **80**, 174406 (2009).
[92] S. Valencia, L. Peña, Z. Konstantinovic, L. Balcells, R. Galceran, D. Schmitz, F. Sandiumenge, M. Casanove, and B. Martínez, Journal of Physics: Condensed Matter **26**, 166001 (2014).
[93] J. Liu, B. J. Kirby, B. Gray, M. Kareev, H. U. Habermeier, G. Cristiani, J. W. Freeland, and J. Chakhalian, Phys. Rev. B **84**, 092506 (2011).
[94] F. A. Cuellar, Y. H. Liu, J. Salafranca, N. Nemes, E. Iborra, G. Sanchez-Santolino, M. Varela, M. G. Hernandez, J. W. Freeland, M. Zhernenkov, M. R. Fitzsimmons, S. Okamoto, S. J. Pennycook, M. Bibes, A. Barthélémy, S. G. E. te Velthuis, Z. Sefrioui, C. Leon, and J. Santamaría, Nature Communications **5**, 4215 (2014).
[95] J. Salafranca and S. Okamoto, Phys. Rev. Lett. **105**, 256804 (2010).
[96] V. Jaccarino and M. Peter, Phys. Rev. Lett. **9**, 290 (1962).
[97] T. J. Liu, J. C. Prestigiacomo, Y. M. Xiong, and P. W. Adams, Phys. Rev. Lett. **109**, 147207 (2012).
[98] J. W. Freeland, J. Chakhalian, A. V. Boris, and J. M. Tonnerre, Phys. Rev. B **81**, 094414 (2010).
[99] B. Nanda, S. Satpathy, and M. S. Springborg, Phys. Rev. Lett. **98**, 216804 (2007).
[100] R. Scherwitzl, S. Gariglio, M. Gabay, P. Zubko, M. Gibert, and J. M. Triscone, Phys. Rev. Lett. **106**, 246403 (2011).
[101] A. T. Lee and M. J. Han, Phys. Rev. B **88**, 035126 (2013).
[102] S. Dong and E. Dagotto, Phys. Rev. B **87**, 195116 (2013).
[103] S. Wu, S. Cybart, D. Yi, J. Parker, R. Ramesh, and R. Dynes, Phys. Rev. Lett. **110**, 067202 (2013).
[104] C. A. F. Vaz, Journal of Physics: Condensed Matter **24**, 333201 (2012).
[105] D. Yi, J. Liu, S. Okamoto, S. Jagannatha, Y.-C. Chen, P. Yu, Y.-H. Chu, E. Arenholz, and R. Ramesh, Phys. Rev. Lett. **111**, 127601 (2013).
[106] A. M. Goldman, Annu. Rev. Mater. Res. **44**, 45 (2014).
[107] C. H. Ahn, M. Di Ventra, J. N. Eckstein, C. D. Frisbie, M. E. Gershenson, A. M. Goldman, I. H. Inoue, J. Mannhart, A. J. Millis, A. F. Morpurgo, D. Natelson, and J.-M. Triscone, Rev. Mod. Phys. **78**, 1185 (2006).
[108] S. V. Kalinin, A. Borisevich, and D. Fong, ACS Nano **6**, 10423 (2012).
[109] J. Jeong, N. Aetukuri, T. Graf, T. D. Schladt, M. G. Samant, and S. S. P. Parkin, Science **339**, 1402 (2013).
[110] J. D. Jorgensen, B. Dabrowski, S. Pei, D. G. Hinks, L. Soderholm, B. Morosin, J. E. Schirber, E. L. Venturini, and D. S. Ginley, Phys. Rev. B **38**, 11337 (1988).
[111] H. Jeen, W. S. Choi, M. D. Biegalski, C. M. Folkman, I.-C. Tung, D. D. Fong, J. W. Freeland, D. Shin, H. Ohta, M. F. Chisholm, and H. N. Lee, Nat Mater **12**, 1057 (2013).
[112] L. Y. Zhu, Y. Liu, F. S. Bergeret, J. E. Pearson, S. G. E. te Velthuis, S. D. Bader, and J. S. Jiang, Phys. Rev. Lett. **110**, 177001 (2013).
[113] C. Visani, J. Tornos, N. M. Nemes, M. Rocci, C. Leon, J. Santamaría, S. G. E. te Velthuis, Y. Liu, A. Hoffmann, J. W. Freeland, M. Garcia-Hernandez, M. R. Fitzsimmons, B. J. Kirby, M. Varela, and S. J. Pennycook, Phys. Rev. B **84**, 060405(R) (2011).
[114] J. W. Seo, W. Prellier, P. Padhan, P. Boullay, J. Y. Kim, H. Lee, C. D. Batista, I. Martin, E. E. M. Chia, T. Wu, B. G. Cho, and C. Panagopoulos, Phys. Rev. Lett. **105**, 167206 (2010).
[115] Y. Liu, S. G. E. te Velthuis, J. W. Freeland, N. J. Tornos, C. Leon, and J. Santamaría, in (2013), p. 17012.
[116] L. Li, C. Richter, J. Mannhart, and R. C. Ashoori, Nat Phys **7**, 762 (2011).
[117] J. S. Lee, Y. W. Xie, H. K. Sato, C. Bell, Y. Hikita, H. Y. Hwang, and C. C. Kao, Nat Mater **12**, 703 (2013).